\documentclass[
aps,%
12pt,%
final,%
notitlepage,%
oneside,%
onecolumn,%
nobibnotes,%
nofootinbib,%
superscriptaddress,%
showpacs,%
centertags]%
{revtex4}
\usepackage[]{graphicx}

\def\gsim{\mathrel{\raise.3ex\hbox{$>$\kern-.75em\lower1ex\hbox{$\sim$}}}}
\def\lsim{\mathrel{\raise.3ex\hbox{$<$\kern-.75em\lower1ex\hbox{$\sim$}}}}

\begin{document}

\title{Electrostatic interaction in plasma with charged Bose condensate.}
\author{\firstname{A.}~\surname{Lepidi}}
\email{lepidi@fe.infn.it}
\affiliation{Universit\`a degli Studi di Ferrara, 
and Istituto Nazionale di Fisica Nucleare Sezione di Ferrara, Italy}

\begin{abstract}
Screening in plasma with Bose-Einstein condensate is studied.
Finite temperature effects are taken into account.
It is shown that, due to condensate effects, the potential has several unusual features. It contains two oscillating terms, one of which is analogous to the fermionic Friedel oscillations in standard QED, and a power law decreasing term. 
In the $T \rightarrow 0$ limit, only one of the oscillating terms survives.
On the whole, any charge impurity is screened more efficiently than in ordinary plasma.
\end{abstract}

\maketitle

\section{Introduction}

It is a well known fact that any interaction may be affected by medium effects.
As an example let us consider the electrostatic potential generated by some charge impurity $Q$, that is described in vacuum by the standard Coulomb potential:
	\begin{eqnarray}
	U(r) = \frac{Q}{4 \pi r}.
	\end{eqnarray}
If the same interaction takes place in a medium, the resulting potential, called Yukawa potential, is exponentially damped, a phenomenon called screening: 
	\begin{eqnarray}
	U(r) = \frac{Q}{4 \pi r} \, e^{-m_D x}.
	\end{eqnarray}
In the last equation $m_D$ is the Debye mass, that depends on the medium features, such as its temperature, the mass and the chemical potential of the composing particles and so on. 
It has been calculated for several media, such as a massive charged fermion gas (e.g. $e^- \, e^+$) and the corresponding boson system ($\phi^- \, \phi^+$) both at finite temperature $T$ and chemical potential $\mu$ \cite{Kapusta:1989tk}. Some examples that can be found in the literature are plasma consisting of:
	\begin{itemize}
	\item Relativistic fermions with $m_F \ll T, \mu_F$:
		$m_{D }^2  = e^2\left( T^2/3 + \mu_F^2 / \pi^2 \right)$.
	\item Non relativistic fermions: 
		$ m_D^2 = e^2 n_F / T$,
		where $ n_F = \frac{\exp (\mu/T)}{\pi^2}\,\int dq q^2 e^{-q^2/2m_FT} $.
	\item Massless scalars without chemical potential:
		$ m_{D }^2  = e^2 T^2/3$.
	\end{itemize}
Screening can be intuitively understood in terms of polarization: the charge impurity polarizes the medium and hence the resulting effective interaction is weaker. 

Surprisingly, till a few years ago, no one has considered what happens when the interaction takes place in a medium with a Bose-Einstein condensate (BEC) component. 

The BEC is a collection of particles which reside all in the same quantum state. Its existence was predicted in 1925 by S. N. Bose and A. Einstein, although it took 70 years to observe it experimentally. 
The first observation was performed in 1995 at the University of Colorado using a gas of rubidium atoms cooled to 170 nK  \cite{Anderson:1995gf}. It was awarded with the 2001 Nobel Prize in Physics. 

To deal with BEC in quantum field theory (QFT) it is convenient to adopt a statistical approach. It is well known that the equilibrium distribution function of a collection of bosons is, up to spin counting factor:
%
	$f_B = (\exp [(E-\mu_B)/T] -1)^{-1} $,
%
where $\mu_B$ is the boson chemical potential, assumed to be smaller than the boson mass $m_B$. One can easily verify that something goes wrong with this distribution function when the boson asymmetry is so large to require a chemical potential larger than the boson mass: the distribution function becomes negative for low momenta, that is physically senseless. 
This is the signal that something happens when the critical value $\mu_B = m_B$ is reached: the system undergoes a phase transition and a BEC is formed. One can easily verify that the equilibrium distribution function of particles and antiparticles for $\mu_B > m_B$ are: 
	\begin{eqnarray}
	\label{Bose_cond_distr}
	f_B = C \delta^{(3)} (\mathbf{p}) + \frac{1}{\exp [(E-m_B)/T] -1}
	\hspace{2cm}
	f_{\bar B} = \frac{1}{\exp [(E + m_B)/T] -1}.
	\end{eqnarray}
The first term in the right hand side of $f_B$, that is proportional to a delta function and has amplitude $C$, represents the condensate.
Using this formalism, the BEC is thus represented as a collection of particles of amplitude $C$, having zero momentum and spatially spread distribution function.

The problem of BEC in QFT is of interest for several reasons. First of all, BECs may be realized in the primordial phases of the universe evolution, in particular around the EW symmetry breaking epoch \cite{Linde:1979pr} as well as in astrophysical systems \cite{Gabadadze:2008mx} and hence it is important to understand what are the consequences of its presence. 
In particular, we analyze in this talk a simple abelian $U(1)$ model, that is interesting for application to white dwarfs astrophysics and serves as a toy model for the subsequent analysis of non abelian theories and application to cosmology. 
Besides the simplicity of the model, we found a rather unusual behavior of the electrostatic potential.

Secondly, elementary charged scalar particles as the ones present in our model characterize several  popular models for physics beyond the standard model, e.g. the class of the SuperSymmetric theories (SUSY), where a boson partner is associated to any fermion of the standard model. These particles may in general form condensates, so our work may be useful as a toy model for understanding possible condensate phenomena in these theories.

\section{The model}

Let us consider an abelian gauge theory containing massive charged fermions and (scalar) bosons: 
	\begin{eqnarray*}
	\label{L_SQED}
	L = -\frac{1}{4} F_{\mu\nu} F^{\mu\nu} - m_B^2 |\phi|^2
	+ |(\partial_{\mu} + i\,e A_{\mu}) \phi |^2
	+ \bar \psi (i \partial \!\!\! / - e A \!\!\! /  - m_F) \psi
	\end{eqnarray*}
and assume finite temperature, $T \neq 0$, global charge neutrality, $Q = 0$, and non vanishing fermion chemical potential, $\mu_F \neq 0$.
As we explained in the Introduction, bosons condense at sufficiently high $\mu_F$, when one needs at least $\mu_B = m_B$ to maintain charge neutrality. 
The BEC amplitude depends on the system temperature, on the fermion chemical potential and on the particle masses \cite{Dolgov:2008pe}: 
	\begin{eqnarray}
	\hspace{-0.2cm}
	C= - 4 \pi
	\int \!\! dq \, q^2
	 \left[  f_B(E_B,m_B,T) - \bar f_B(E_B,-m_B,T)
	 - 2 f_F(E_F,\mu_F,T) + 2 \bar f_F(E_F,\mu_F,T)
	\right]
	\end{eqnarray}
The electrostatic potential $U(r)$ is calculated from the time-time component of the photon polarization tensor $\Pi^{\mu\nu}$, that enters the photon equation of motion as:
	\begin{eqnarray}
	\left[ k^\rho k_\rho g^{\mu\nu} - k^\mu k^\nu + 
	\Pi^{\mu\nu}(k)
	\right] A_\nu (k) 
	= \mathcal{J}^\mu (k).
	\end{eqnarray}
$\Pi^{\mu\nu}$ contains all the informations concerning the photon interaction with the other particles.
Technically, we calculated the photon Green function using standard perturbative techniques up to the second order in the coupling constant $e$. The thermal effects were taken into account by calculating the average of the propagator over the thermal bath \cite{Dolgov:2008pe}.
As a result, the boson - see Eqs.(\ref{Bose_cond_distr}) - and fermion distribution functions:
	\begin{eqnarray}
	\label{}
	f_F = \frac{1}{\exp [(E - \mu_F)/T] +1}
	\hspace{2cm}
	f_{\bar F} = \frac{1}{\exp [(E + \mu_F)/T] +1},
	\end{eqnarray}
enter the calculation of $\Pi^{00}$:
	\begin{eqnarray}
	\Pi_{00}  
	(k) &=&
	e^2 \, 
	\int \frac{d^3 q}{(2\pi)^3 E}   C \, \delta^{3}(q)
	\left( 1 + \frac{4 E^2}{k^2} \right) 
	\cr\cr
	&+& 
	\frac{e^2}{2\pi^2} \int_0^\infty \frac{dq\,q^2}{E} 
	\left[f_F(E_F,\mu_F,T)+ \bar f_F(E_F,-\mu_F,T)\right] 	
	\left[2+\frac{ (4E_F^2 - k^2)}{2 k q} \ln \bigg| \frac{2q +k}{2q - k} \bigg| \right]
	\cr\cr
	&+& 
	\frac{e^2}{2\pi^2} \int_0^\infty  \frac{dq \,q^2}{E} 
	\left[ f_B(E_B,m_B,T) + \bar f_B(E_B,-m_B,T)\right] 	
	\left[1+   \frac{ E^2_B}{ k q} \ln \bigg| \frac{2q +k}{2q - k} \bigg| \right]
	\end{eqnarray}
Once that $\Pi^{00}$ is known, the electrostatic potential $U(r)$ can be calculated as:
	\begin{eqnarray}	
	U(r) = Q \int \frac{d^3 k}{(2\pi)^3} \frac{\exp (i {\bf k r)}}{k^2 - \Pi_{00} (k)} 
	=
	\frac{Q}{2\pi^2}\int_0^\infty \frac{dk k^2}{k^2 - \Pi_{00} (k)}\,\frac{\sin kr}{kr}.
	\end{eqnarray}
When the BEC is absent, $\Pi_{00}$ is independent of the photon momentum $k$, hence the standard Yukawa screening comes from poles at purely imaginary $k$. In this case the Debye mass is simply $\Pi_{00} = -m_D^2$.
When a BEC is present, the situation is more complicated because $\Pi_{00}$ is k-dependent and infrared singular \cite{Dolgov:2008pe}:
	\begin{eqnarray}
	\hspace{2.5cm} \Pi_{00} &=& - e^2 \left[ \left(
	m_0^2 + \frac{m_1^3}{k} + \frac{m_2^4}{k^2}
	\right)\right]
	\cr\cr
	m^2_0 = \left( \frac{2T^2}{3} \right)+ \frac{C}{(2\pi)^3m_B} \hspace{1cm}
	&& m_1^3 =  \frac{m_B^2 T}{2} \hspace{2.2cm}
	m_2^4 = \frac{4Cm_B}{(2\pi)^3}
	\end{eqnarray}
As a result of the peculiar form of the photon polarization tensor, the electrostatic potential is a rather complicated function \cite{Dolgov:2008pe,Dolgov:2009yt}:
	\begin{eqnarray}
	\label{U_tot}
	U(r) &=& 
	- \frac{Q}{4 \pi }
	\, \frac{\exp ( - \sqrt{e/2} m_2 r) \cos (\sqrt{e/2} m_2 r)}{r} 
	- \frac{12 Q \, m_1^3}{\pi^2m_2^8 } \, \frac{1}{e^2} \, \frac{1}{r^6} \cr\cr
	&& +  \frac{8 (2 \pi)^6 Q \, T^3 m_B^2}{C^2}\,
	\frac{1}{e^2} \, \frac{1}{r^2}
	\exp \left(-2\sqrt{2\pi m_B T} r\right)\,\cos\left(2\sqrt{2\pi m_B T} r\right).
	\end{eqnarray}
The first oscillating term comes from pole contributions analogous to the standard Debye screening. 
The same term without the temperature corrections was found in \cite{Gabadadze:2008pj} by different technique.
One can note that in this case the "Debye mass" is non analytic in the coupling constant $e$, being it proportional to $\sqrt{e}$
\footnote{Please note that the localization of the pole at $\sqrt{e/2} m_2$ is exact only in the $T\rightarrow 0$ limit. Otherwise the pole is shifted and the analytic calculation of a generic solution for any value of the parameters becomes impossible. For more details see \cite{Dolgov:2009yt}.}. 
This term is the only one surviving in the $T \rightarrow 0$ limit. 
The second term is a non-perturbative contribution rapidly decreasing as $r^{-6}$. In the limit $\mu_B = m_B$, $C=0$, it is modified to \cite{Dolgov:2009yt}:
	\begin{eqnarray}
	\label{}
	- \frac{2Q}{\pi^2 m_B^2 T} 
	\, \frac{1}{e^2} \, \frac{1}{r^4} .
	\end{eqnarray}
We can conclude that the condensate always manifests itself by a strong asymptotic increase of screening.
The third term in Eq.(\ref{U_tot}) is a finite temperature effect. It is analogous to Friedel oscillations in QED, which on the contrary are present at $T=0$ and exponentially fade away at finite temperature - for a complete discussion on this point see \cite{Dolgov:2009yt}.

\section{Conclusion}

The BEC is a purely quantum state of matter having very interesting features. 
It can be realized in laboratory, a field of research that became very active in the last 15 years as well as in cosmological and astrophysical environments.
For these last applications it is convenient to adopt a quantum field theory approach, even though the effects of BEC on quantum field theories are still not completely explored.

This talk is based on the papers \cite{Dolgov:2008pe,Dolgov:2009yt}, where we considered a simple abelian $U(1)$ gauge theory including a BEC and analyzed the effects of the condensate on the interacting potential. 
In particular, we studied in detail the electrostatic interaction and found a very peculiar potential, that oscillates with distance and temperature, it is non analytic in the coupling constant $e$ and screens very efficiently any charge impurity. 

\noindent 
\\ \\
{\bf Acknowledgments} 
The author is thankful to A. Dolgov for careful reading of the manuscript and useful suggestions.

%
%


\end{document}